\documentclass[twocolumn,prl,superscriptaddress]{revtex4}
\usepackage[dvips]{graphicx}

\begin{document}

\def\x{\vec{x}}
\def\v{\vec{v}}

\title{Network formation of tissue cells via preferential attraction to elongated structures}

\author{Andras  Szabo}
\affiliation{
Dept of Biological Physics, Eotvos University, Budapest, Hungary
}

\author{Erica D. Perryn}
\affiliation{
Dept of Anatomy \& Cell Biology, KU Med. Ctr., Kansas City, KS
}

\author{Andras  Czirok}
\affiliation{
Dept of Biological Physics, Eotvos University, Budapest, Hungary
}
\affiliation{
Dept of Anatomy \& Cell Biology, KU Med. Ctr., Kansas City, KS
}
\email{czirok@biol-phys.elte.hu}

\date{\today}

\begin{abstract}
Vascular and non-vascular cells often form an interconnected network in vitro,
similar to the early vascular bed of warm blooded embryos. Our 
time-lapse recordings show that the network forms by extending sprouts,
i.e., multicellular linear segments. To explain the emergence of such
structures, we propose a simple model of preferential attraction to
stretched cells. Numerical simulations reveal that the model evolves 
into a quasi-stationary pattern containing linear segments, which 
interconnect above the critical volume fraction of 0.2. In the quasi-stationary state the 
generation of new branches offset the coarsening driven by surface tension. In
agreement with empirical data, the characteristic size of the resulting
polygonal pattern is density-independent within a wide range of volume fractions.
\end{abstract}

\pacs{87.17.Jj, 87.18.Bb, 87.18.Ed, 87.18.Hf}
\maketitle

\begin{figure}
\includegraphics[width=90mm]{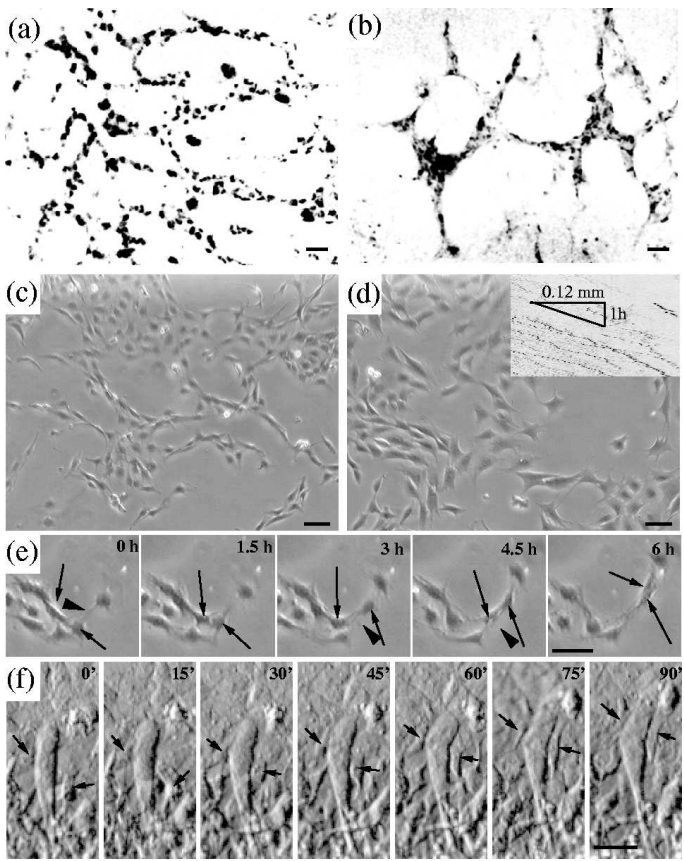}
\caption{ \label{fig1}
Tissue cell networks. (a) Vascular endothelial cells during vasculogenesis, 
visualized in a bird embryo according to \cite{CRRL02}. (b) Mouse vascular 
endothelial cells within an in vitro explant \cite{Drake00a}, after 24 h of 
culturing.
(c) Astroglia-related rat C6 cells 
and (d) muscle-related mouse C2C12 cells cultured on a rigid tissue culture 
substrate as described in \cite{CSMV98}. The inset depicts the motion of latex 
beads at the tissue culture surface. 
To measure convection currents, 10
$\mu$m wide and 250 $\mu$m long rectangles were selected in the images, directed
parallel to the flow. For each frame, pixel intensities were averaged across
the rectangle, resulting in the horizontal lines of the inset.
(e) C6 cells (arrows) migrate along extended
projections of adjacent cells (arrowheads). 
(f) Endothelial cells (arrows)
move along a vascular sprout, visualized in the mouse explant by differential 
interference contrast microscopy. 
}
\end{figure}

Embryogenesis, the shaping of tissues and organs, is a long standing challenge 
of science. Recent application of ideas
originating from non-equilibrium statistical physics has proved to be fruitful
for understanding emergent properties such as pattern formation, or biomolecular 
network function during embryo development \cite{Forgacs05}.
Formation of the primordial vascular bed of warm blooded vertebrates is
arguably one of the best examples for an emergent phenomenon in embryonic
development \cite{RF95}. During vasculogenesis, well before the onset of circulation,
hundreds of essentially identical vascular endothelial cells create a polygonal
network within a simple, sheet-like anatomical environment
(Fig.~\ref{fig1}a,b).  The network is highly variable among individuals within
certain statistical constraints \cite{LaRue03b}, relatively uniform in morphology, and unlike
insect segmentation, there is no evidence for direct genetic control of
vascular segment positions.

The formation of linear segments and their interconnected network is not
restricted to vascular endothelial cells \cite{VLS95}. In particular, we demonstrate in
Fig.~\ref{fig1} (c) and (d) that non-vascular, glia- or muscle-related cells
also exhibit linear structures when grown on a rigid plastic tissue culture
substrate with continuously shaken culture medium. Depending on the cell
density, the linear segments can merge and form an irregular network.

Recent in vivo observations of vasculogenesis indicated that early 
vascular network
formation includes {\em sprouting}, the extension of linear segments containing multiple cells \cite{Rupp04}.
This process is markedly different from the gradual coarsening of an initially
uniform density field, and its possible arrest, characteristic for colloid 
gels (see, e.g. \cite{Foffi05}). To
determine the mechanism of sprout formation, we recorded the temporal
development of the structures shown in Fig.~\ref{fig1}(b-d) with computer controlled
microscopy \cite{CRRL02}. In vitro culture conditions yield sufficiently high resolution to trace the motion of
individual cells during the patterning process. As shown in
Fig.~\ref{fig1}(e,f), sprout expansion involves cell motility guided by
adjacent projections of other cells or elongated multicellular structures.

A number of theoretical models were formulated to understand the
self-organization of vascular networks.  The {\em chemo-mechanical} mechanism
assumed cells to exert mechanical stress on the underlying substrate, and the
resulting stress to guide cell motility \cite{Murray98,Murray2}.  A recent model of
Gamba, Serini et al assumed {\em chemoattractant} signaling 
\cite{Gamba03,Serini03}. While the suggested chemoattractant, VEGF$_{165}$, is
unlikely to behave in the predicted way (see the discussion in \cite{Merks06}),
patterning guided by an unspecified chemoattractant continues to serve as the
basis of biologically plausible models including cell adhesiveness and finite
cell size \cite{Merks06}.  

Both the chemo-mechanical and chemoattractant mechanisms may be biologically 
relevant under certain circumstances. We argue, however, that neither can
explain sprout formation seen in Fig.~\ref{fig1}.
In both models the pattern were reported to arise in a gradual coarsening 
process in which segments are eliminated and small holes are replaced with 
larger ones. Moreover, neither mechanism is expected to operate within our in 
vitro experimental setup: The rigid substrate excludes the chemo-mechanical
mechanism. A specific chemotactic response is
unlikely to be shared by such a variety of cell types.
Finally, convection currents in the culture medium,  generated by  
temperature inhomogeneities within the incubator and the vibrations of
microscope stage motion, are expected to hamper the 
maintenance of concentration gradients, or impose a strong 
directional bias upon the chemotaxis-related cell movements. 

To measure
the convection currents close to the culture surface, we immersed 
0.5 $\mu$m diameter latex beads (Sigma) into the medium.
Bead motion was recorded within a 20 $\mu$m thick volume above the culture 
surface, delimited by the field depth of the 10X 
microscope objective.  As a representative sample (Fig.~1d, inset)
demonstrates, in our experimental setup convection currents were 
sustained for hours with speeds exceeding 100 $\mu$m/h, an order of 
magnitude larger than the median cell speed.

On the basis of our empirical observations, in this letter we propose that
cellular sprouting employs a quite generic mechanism, the {\em preferential
attraction to elongated structures}.  While the molecular basis of such a
behavior is unknown, one may conjecture two biologically feasible scenarios.
(i) The tendency of cells to align with one another was analyzed in detail in
Myxobacteria (see, e.g. \cite{Pelling06}), and similar mechanisms may also
operate in animal cells. (ii) Cells in elongated structures are assumed to 
be under mechanical tension, and strained cells can have a stiffer cytoskeleton
\cite{Xu00}. Cells are able to respond to variations in extracellular matrix
stiffness \cite{Gray03}, and an analogous mechanotaxis utilizing cell-cell 
contacts is also feasible. To assess the collective behavior of cells
exhibiting the proposed preferential attraction property, we studied the
following simple model in which individual cells are represented as particles. 

\begin{figure}
\includegraphics[width=90mm]{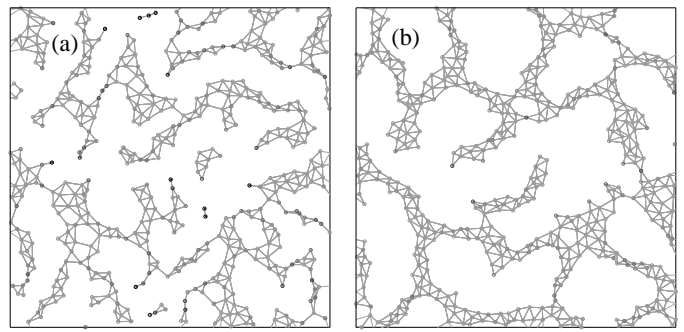}
\caption{ \label{fig2}
Network formation in the model. Randomly placed $N=500$ particles assemble into
linear structures, detectable already within 30' (a). At sufficiently high
particle density a characteristic 
pattern size develops in five hours (b) with a combination of 
sprouting (branch extension) and coarsenening (merger of adjacent branches).
Connected dots represent Voronoi neighbor particles. Darkening gray levels
indicate increasing local anisotropy. The simulation covered an area of 
$L=0.7$mm. 
}
\end{figure}

Cell motility is often approximated as a persistent diffusion process 
\cite{Stokes91,Selmeczi05}, where the velocity $\v_k$ of cell $k$ is 
described by the Langevin equation
\begin{equation}
{d \v_k\over dt}=-\v_k/\tau+\sqrt{D}\xi_k+\vec{M}_k
\label{eq1}
\end{equation}
where $\tau$ is the persistence time of motion, $D$ is a diffusion
parameter, and $\xi$ is an uncorrelated white noise: $\langle \xi \rangle=0$ and 
$\langle \xi_k(t)\xi_l(t')\rangle=\delta_{kl}\delta(t-t')$. 
Term $\vec{M}$ is a deterministic bias, representing interaction with
the environment (specified below). 
While Eq.(\ref{eq1}) describes the motion of a Brownian particle at finite
temperatures, animal cell motility is driven by complicated molecular 
machinery, and it is {\em not} thermal fluctuation driven. Thus, parameters
$\tau$ and $D$ depend substantially on cell type and molecular state.
Measurements performed with non-interacting endothelial cells and fibroblasts
resulted $\tau$ and $D$ values in the $0.1-5$ h and 100-2000 $\mu$m$^2$/h$^3$ 
range, respectively \cite{Dunn83,Stokes91}.

Interactions among mobile agents are usually modelled as a sum of pair
interactions \cite{Helbing01}. In this spirit, we factor $\vec{M}$ into
\begin{equation}
\vec{M}_k = \sum_{\{j\}} {\x_j-\x_k\over d_{kj}}\left[f_1(d_{kj}) + w_jf_2 (d_{kj})\right],
\label{eq2}
\end{equation}
where the sum is taken over the Voronoi neighbors of particle $k$, and
$d_{kj}=\vert \x_j-\x_k \vert$.  
The soft-core repulsion $f_1$ ensures that model cells are impenetrable. The
range of repulsion is the size $R_1$ of the organelle-packed region around the
cell nucleus. The preferential attraction response is incorporated in the $f_2$
term. Cells are expected to explore their surroundings with protrusions, and
respond when protrusions contact elongated structures. Filopodia typically extend
from $R_2$, the cell surface $(R_1\leq R_2)$, to a maximal distance of $R$.
Thus, $f_1(d)=0$ for $d>R_1$ and $f_2(d)=0$ for $d<R_2$ and $d>R$. 
Based on Fig.~\ref{fig1} we estimate $R_1=10 \mu$m, 
$R_2=30 \mu$m and $R=40 \mu$m. These values, however, can vary by at least a 
factor of two, depending on the cell types and experimental conditions.

For the sake of simplicity, cell shape is not explicitly resolved in the model. 
Therefore, elongation or local anisotropy is inferred from the
configuration of particles. To represent an attraction to cells within anisotropic 
structures, the weights $w_k$ are constructed as
\begin{equation}
w_k= {1\over n_k}\left|\sum_{\{j :d_{jk}<R\}}e^{2i\varphi_{jk}}\right|^2 
\end{equation}
where the sum is taken over all $n_k$ particles that are within a circle of
radius $R$ around particle $k$. The angle between $\x_k-\x_j$ and an 
arbitrary reference
direction is denoted by $\varphi_{jk}$. Thus, $w=0$ for particles in an
isotropic environment and $w=1$ for particles in a highly elongated, linear
configuration.  Non-uniform weights result in asymmetric pair-interactions,
which is feasible 
as $\vec{M}$ represents a bias in cell activity instead of physical forces.

Eqs (1)-(3) were studied by Euler integration with $0.05$h long time steps within a rectangular area of size $L$.
Random initial and periodic boundary conditions were applied.  Parameter values
$\tau=0.5$h and $D=100 \mu$m$^2$/h$^3$ were chosen to represent cell motility.
There is little empirical guidance on the choice of functions $f_1$ and $f_2$.
We tested (i) a linear form with $R_1=R_2=15\mu$m as $f_1(d)+f_2(d)\sim d-R_1$
and (ii) a Hertz-type repulsion $f_1(d)=-A(R_1-d)^n$ with a
zonal, distance-independent attraction $f_2(d)=B$ for $R_2<d<R$, where the
Hertz exponent $n$, $A$ and $B$ are parameters.  Irrespective of the 
functional forms (i) and (ii), as well as for both $n=1$ and $n=1.5$ 
a parameter regime exists in which sprout-like linear structures
form during a biologically feasible time period. 
In Fig 2. and subsequently we demonstrate simulation results of
model (ii) with parameters $n=1$, $A=160$ h$^{-2}$ and 
$B=130$ $\mu$m/h$^2$. 
These values represent a strong response to external cues:
the ratio of the directed and random velocity components is $B\tau/\sqrt{D\tau}\approx3$. As a comparison, a similar measure for chemotactic response of
endothelial cells was found larger than one \cite{Stokes91}.

\begin{figure}
\includegraphics[width=90mm]{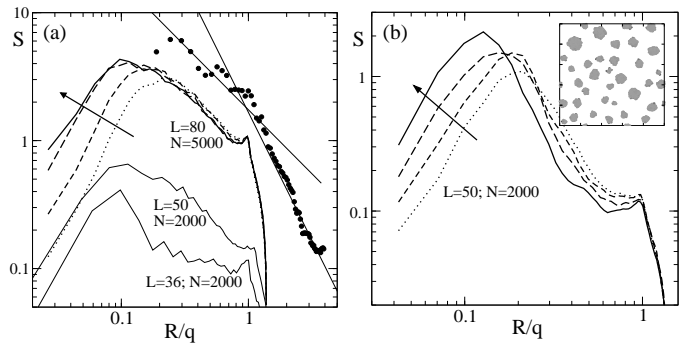}
\caption{ \label{fig3}
Power spectrum $S(q)$ of the particle configurations, at various time points and
model parameters. The dotted, dashed and solid curves indicated by the arrows
were obtained at $t=$25, 75, 125 and $500$h, respectively. 
The asymmetric model (a) reaches a steady state while the symmetric model (b) 
continues to coarsen into droplets (inset).
The shifted curves in (a) demonstrate that $q_*$ is independent of the 
system size and cell density in the stationary phase.
For low cell densities the peak falls off as $1/q$, in good agreement with 
data obtained for Fig.~\ref{fig1}a (filled symbols). The solid lines 
represent power-law decays with exponents $-1$ and $-2$.
} 
\end{figure}

The time development of the system was followed by the
ensemble-averaged binned power spectrum $S(q)=\langle \vert \sum_j \exp(-2\pi
i\vec{q}\x_j) \vert^2 \rangle$, where each component of $\vec{q}$ is an integer
multiple of $1/L$.  The $\langle ... \rangle$ average is taken over
configurations obtained with different noise realizations and over all scatter
vectors $\vec{q}$ of length $q$.  As Fig.~3a demonstrates, after an initial
coarsening regime the pattern reaches a quasi-stationary state in which the
generation of new branches balances the elimination of holes. The spectra
exhibits two peaks; one at $q\approx1/R$ reflecting the typical distance of adjacent
particles, and another characterizing the pattern size at $q_*=1/\ell_*$. In
the quasi-stationary regime the characteristic length $\ell_*/R\approx 10$
is independent of the system size, and only weakly depends on the particle
density.  

The insensitivity of $\ell_*$ on particle density is in good 
agreement with the somewhat limited morphometric data available for the 
vasculature of quail embryos (see Fig.~3 of \cite{LaRue03b}), where 
$R\approx20\mu$m. LaRue
et al characterized the vasculature with the average diameter of avascular areas $M_A$
and the average width of vascular segments $M_V$. From these two quantities,
we estimate the characteristic pattern size as $\ell_*\approx M_A + M_V$
and the volume fraction as $\sigma\approx1-M_A^2/\ell_*^2$. The five 
comparable data points cover volume fractions in the $0.2\leq \sigma \leq 0.9$ 
range. The characteristic pattern size is $\ell_*=80\pm15$ $\mu$m with no 
obvious correlation between $\ell_*$ and $\sigma$.

\begin{figure}
\includegraphics[width=90mm]{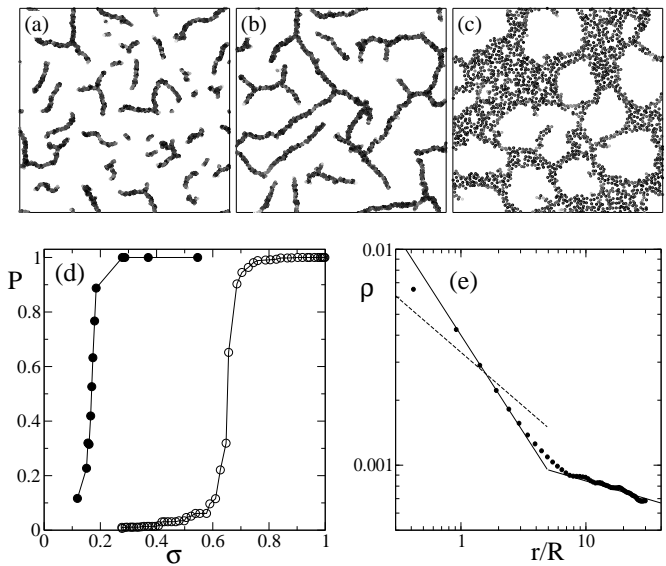}
\caption{ \label{fig4}
Network connectivity depends on particle density. Configurations of $N=2000$
particles are shown in the stationary phase for various values of $N/L^2$ 
particle density: 0.31 (a), 0.44 (b), and 1.54 (c).
The maximal cluster size (d) shows a percolation transition at 
a volume fraction below 0.2 (filled symbols), much less than the critical volume 
fraction of randomly placed disks (open symbols). The density-density 
autocorrelation of the critical cluster (e) is biphasic, and well approximated by 
$\varrho(r)\sim r^{-0.9}$ for $r<r_c$ and $\varrho(r)\sim r^{-0.2}$ for $r>r_c$ (solid lines). For comparison, the dashed line represent $\varrho(r)\sim r^{-0.5}$.
}
\end{figure}

As Fig.~4 demonstrates, the connectivity of the pattern depends on the particle density. We
characterized the percolation transition by calculating the relative size of
the largest interconnected cluster, $P$, and the volume fraction $\sigma$ by
treating each particle as a disk of radius $R$ and calculating the net space 
coverage of the configurations.  The percolation threshold is at
volume fraction $\sigma\approx0.2$: similar to the value reported in \cite{Gamba03}, and substantially smaller than 0.67, the
critical volume fraction for randomly placed overlapping disks \cite{VK81}.

To compare our model to that of \cite{Gamba03}, we analyze the structure of the critical cluster by
calculating its mean density $\varrho$ as a function of radius $r$. Above the
$r/R \approx 1$ lower cut-off length $\varrho(r)$ is well approximated by a
biphasic curve: $\varrho(r)\sim r^{-0.9\pm0.05}$ for $1<r<r_c$ and
$\varrho(r)\sim r^{-0.17\pm0.1}$ for $r_c<r$. 
The crossover length
$r_c\approx6R$ is comparable with $\ell_*$, the characteristic pattern size in
the percolating regime. In agreement,
for a wide range of particle densities, which includes a substantial 
regime above the percolation threshold, the $S(q)$ at $q_*$ falls off to 
increasing $q$ as $S(q) \sim q^{-1}$ (Fig. 3a).  
While the obtained exponents are subject to finite size effects due to the 
limited scaling regime, both $\varrho(r)$ and $S(q)$ is more compatible with
linear structures below $r_c\approx\ell_*$, rather than the
fractal-like behavior $\varrho(r)\sim r^{-0.5}$ seen in the chemoattractant
model.

In conclusion, we demonstrate that a preferential attraction to elongated cells
can be sufficient to explain the abundance of network-like structures in cell
cultures and that it is likely to be an important component of vasculogenic 
patterning. Network formation of mobile agents with spatially limited 
interaction range is a fundamental problem also occurring in technological 
fields: the establishment of a self-organized communicating network between 
mobile robots is one recent example \cite{Glauche03}. We expect that a similar
sprouting mechanism in a system of self-propelled agents can create adaptive 
networks at low volume fractions.

\begin{acknowledgments}
The authors thank E. M\'ehes for his help in the in vitro cell culture 
experiments; the anonymous referees for their comments; C.D. Little and T. Vicsek for valuable discussions. This
work was supported by grants 0535245N, 0410084Z of the American Heart 
Association, HL068855 of NIH, T047055 of the Hungarian Research Fund and
a grant from the G. Harold \& Leila Y. Mathers Charitable Foundation.
\end{acknowledgments}

\end{document}